\documentclass{svproc}
\usepackage{type1cm}
\usepackage{subeqnarray}
\usepackage{graphicx}
\usepackage{footmisc}

\usepackage{tabularx}
\usepackage{booktabs}

\usepackage{amsmath, amssymb, amsfonts,mathtools} 
\usepackage{units} 
\usepackage{eurosym}
\usepackage{csquotes} 
\usepackage[nocomma, short]{optidef}
\newcommand{\R}{\mathbb{R}}

\usepackage{url}

\begin{document}
\mainmatter              
\title{A Quantum Computing Approach for the Unit Commitment Problem}
\titlerunning{QC Approach for the UCP - Accepted at OR Proceedings 2022}  
%
\author{Pascal Halffmann\inst{1} \and Patrick Holzer\inst{1} \and Kai Plociennik\inst{2} \and Michael Trebing\inst{1}}
\authorrunning{Pascal Halffmann et al. - Accepted at OR Proceedings 2022} 
%
\tocauthor{Pascal Halffmann, Patrick Holzer, Kai Plociennik, and Michael Trebing}
\institute{Department of Financial Mathematics, Fraunhofer Institute for Industrial Mathematics, Am Fraunhofer Platz 1, 67663 Kaiserslautern, Germany,\\
\email{pascal.halffmann@itwm.fraunhofer.de},\\ 
\and 
Competence Center High Performance Computing, Fraunhofer Institute for Industrial Mathematics, Am Fraunhofer Platz 1, 67663 Kaiserslautern, Germany}

\maketitle              

\begin{abstract}
Planning energy production is a challenging task due to its cost-sensitivity, fast-moving energy markets, uncertainties in demand, and technical constraints of power plants. Thus, more complex models of this so-called \emph{unit commitment problem (UCP)} have to be solved more rapidly, a task that probably can be solved more efficiently via quantum computing. In this article, we model a UCP with minimum running and idle times as a quadratic unconstrained optimization problem to solve it on quantum computing hardware. First experiments confirm the advantages of our formulation in terms of qubit usage and connectivity and most importantly solution quality.
\keywords{quantum computing, unit commitment problem, quadratic programming, energy planning\\ \textbf{Accepted at OR Proceedings 2022}}
\end{abstract}
\section{Introduction}\label{sec:intro}
An energy supply that is both stable and environmentally sustainable is vital for viable economic growth and social welfare. However, with the recent substantial increases in energy market prices and price fluctuations, energy generation is a highly cost-sensitive field comprising a complex system of power units and grids. The complexity of the so-called \emph{unit commitment problem (UCP)} presents challenges  when it comes to planning energy generation due to technical constraints regarding power units and the power grid -– a task that,  with increasing reliance on renewable energy, is exacerbated further by weather-induced uncertainties. In conclusion, there is a real incentive to calculate the optimal solution to a realistic UCP model in a short period of time. However, solvers on classical computers often cannot accomplish this task, especially when uncertainties from renewable energy supply are considered. 

In contrast, \emph{quantum computing (QC)} provides a new computational paradigm based on counterintuitive phenomena in quantum mechanics, such as considering all possible solutions at once in a state of superposition. This has the potential of rapidly solving various tasks, from optimization problems to simulations and even communication. Admittedly, in the current, so called \emph{noise intermediate scale quantum (NISQ)} era, only few quantum computers are available with a limited amount of qubits, a rather restricted number of connections between them, and low fault-tolerance. Suitable use cases have just been discovered and algorithms are in development. Nevertheless, QC is a promising technology of the near-term future and has already achieved some remarkable results for example in chemistry \cite{qcc} and finance \cite{qcf}. Therefore, the question arises whether QC will eventually be able to provide solutions to the UCP faster and, due to its inherent capability of coping with uncertain and stochastic parameters, has advantages when considering e.g. uncertain supply from renewable energies.

\subsection{Previous Work}\label{subsec:prevwork}
The unit commitment problem has been studied extensively in the last decades. There exist numerous versions with varying complexity and purpose. We refer to \cite{uc2} for an overview. It is beyond the scope of this article to provide a full literature overview on the extensively published topic of quantum computing. So far, only two contributions try to solve the UCP via quantum computing: In \cite{qc1}, a UCP problem with quadratic cost function, demand satisfaction and minimum and maximum power generation is discussed. They use a generic method to transform this model to a quadratic unconstrained problem. A computational study with three to twelve power units shows that while D-Wave returns near-optimal solutions, the computation via Gurobi is 30000 times faster for the largest instance. Recently, the authors of \cite{qc2} considered a quantum computing approach for a distributed UCP, where power units are concentrated in a connected hub which allows a decomposition into subproblems. They apply a quantum version of the decomposition and coordination alternate direction method of multipliers (ADMM) to this problem. While they report that demand satisfaction, minimum running and idle times, ramping, and power grid constraints among others are considered in the model using the same generic method for the problem transformation as before, a full formulation of their problem is not given.

\subsection{Our Contribution}\label{subsec:ourcon}
In order to solve an optimization problem via a quantum computer it can be transformed into a so-called \emph{Ising Hamiltonian}, which is an operator measuring the total energy of a physical system. This operator has a one-to-one correspondence to a QUBO. Generically, a linear optimization problem with constraints is transformed into a QUBO by transforming inequality constraints to equality constraints via slack variables and adding equality constraints as quadratic penalty terms to the objective function. At last, continuous and integer variables are encoded and replaced by binary variables. Clearly, this method is not problem-specific and has some drawbacks: Each slack variable (or its binary encoding, if it is non-binary) requires additional qubits. The quadratic penalty term results in quadratic terms using every combination of variables used for this penalty term, thus, in an all-to-all connection between the corresponding qubits. Current NISQ devices have a limited number of qubits with restricted interconnectivity and in particular no all-to-all connectivity, which requires additional qubits to facilitate the connection. It is therefore crucial to avoid introducing large numbers of qubits with many interactions. Even in the future, the degree of interconnectivity between qubits can be expected to present a limiting factor.

We propose a novel formulation for a unit commitment problem modeling minimal variable and starting costs, demand satisfaction as well as minimum running and idle times as a QUBO problem. This formulation is explicitly designed for reducing the number of and the connectivity between qubits, while ensuring that all constraints are satisfied even though they are transformed to penalty terms. This has been especially achieved by avoiding slack variables and, largely, squared sums of variables in our method. As pointed out, quantum computing is still in its infancy. We do not expect that solving a UCP with current QC hardware and algorithms can cope with real-world sized problem instances or is competitive against classical approaches. We merely present a proof-of-concept for an alternative direction: given the current rapid development of QC and its present successes, our model may provide faster or better solving of the UCP in the future, especially for large instances with e.g. uncertain demand.

The remainder of this article is organized as follows: In Section~\ref{sec:classicform}, we formally introduce the unit commitment problem and present a formulation as mixed-integer linear problem. In the next section, our formulation as quadratic unconstrained binary problem is given. An illustrative example of our formulation, together with a comparison to resource requirements for the generic method, and solving the model both via a quantum computer simulator and a quantum annealer, follows in Section~\ref{sec:example}. We conclude this article with an outlook on future directions of research. 

\section{Linear Formulation of the Unit Commitment Problem}\label{sec:classicform}

In this section we present the classical mixed-integer linear formulation of the unit commitment problem. In general, the UCP  deals with finding a cost-minimal operation schedule for a set of thermal units $i\in\{1,\ldots,N\}$ to meet a given demand for electricity over a distinct set of time steps $t\in\{1,\ldots T\}$. Further, technical properties of both the thermal units and the underlying power grid have to be respected. Take note that there does not exist one single UCP, but several variants are present in the literature. For an overview on different formulations, we refer to \cite{uc1}. In the remainder of this article, we propose a model with the following assumptions. Each thermal unit $i$ has the following properties: linear, production dependent costs $varcost_i$ and fixed costs $startcost_i$ for starting the unit, minimum and maximum power generation output, $mingen_i$, $maxgen_i$, and minimum running time and minimum idle time, $minup_i$, $mindown_i$. At each time step $t$ the residual demand $rd_t$, demand minus supply by renewable energies plus spinning reserve, has to be met. Limitations due to the power grid are omitted. This type of unit commitment problem is commonly modeled as follows:

\scriptsize
\begin{mini*}[2]
	{}{\sum_{t=1}^T \sum_{i=1}^I \left(varcost_i\cdot gen_{t,i} + startcost_i\cdot start_{t,i}\right) }{}{}
	\addConstraint{\sum_i gen_{t,i}}{=rd_t,}{\forall~t=1,\ldots,T,}
	\addConstraint{on_{t,i}\cdot mingen_i}{\leq gen_{t,i},}{\forall~t=1,\ldots,T,~i=1,\ldots,I,}
	\addConstraint{on_{t,i}\cdot maxgen_i}{\geq gen_{t,i},}{\forall~t=1,\ldots,T,~i=1,\ldots,I}
	\addConstraint{on_{t,i}-on_{t-1,i}}{\leq start_{t,i},}{\forall~t=1,\ldots,T,~i=1,\ldots,I,}
	\addConstraint{\sum_{\tau=t}^{t-1+minup_i}on_{\tau,i}}{\geq start_{t,i}\cdot minup_i,}{\forall~t=1,\ldots,T,~i=1,\ldots,I,}
	\addConstraint{\sum_{\tau=t+1-mindown_i}^{t}start_{\tau,i}}{\leq 1-on_{t-mindown_i,i},}{\forall~t=1,\ldots,T,~i=1,\ldots,I,}
	\addConstraint{on_{t,i}, start_{t,i}}{\in\mathbb{B}}{\forall~t=1,\ldots,T,~i=1,\ldots,I,}
	\addConstraint{gen_{t,i}}{\in\mathbb{R}_{\geq 0}}{\forall~t=1,\ldots,T,~i=1,\ldots,I.}
\end{mini*}

\normalsize
Here we have three sets of decision variables $on_{t,i}$, $gen_{t,i}$ and $start_{t,i}$, $t\in\{1,\ldots T\}, i\in\{1,\ldots,N\}$. The binary variable $on_{t,i}$ observes whether unit $i$ is running at time $t$. With $gen_{t,i}\in\R_{\geq 0}$ we denote the power generated by unit $i$ at time step $t$. The variables $start_{t,i}\in [0,1]$ track the starting of power units.


\section{Quadratic Unconstrained Binary Formulation of the Unit Commitment Problem}\label{sec:quboform}

In quantum physics, the \emph{Hamiltonian}, an operator corresponding to the total energy of the system it refers to, is used to calculate the energy-minimal state of that system. One of the most famous Hamiltonians is the \emph{Ising Hamiltonian} describing the energy of a solid in a ferromagnetic field using $n$ spins $s_i={\pm 1}$:

\scriptsize
\begin{equation*}
	H(s_1,\ldots,s_n) = -\sum_{i<j} J_{ij}\cdot s_i \cdot s_j - \sum_{i=1}^N h_i\cdot s_i.
\end{equation*}
\normalsize
Due to its close resemblance to a quadratic unconstrained binary problem\\ $\min_{x\in\mathbb{B}^n} x^\top Q x$,
optimization problems are transformed to QUBOs and then to Ising Hamiltonians in order to solve these problems on a quantum computer, see \cite{glover} for methods for the transformation. Spins can be transformed to binary variables via $x_i=\nicefrac{(s_i+1)}{2}$. We concisely state the UCP as QUBO as follows:

\scriptsize
\begin{mini*}[2]
	{}{\sum_{t=1}^T \sum_{i=1}^I \left(varcost_i\cdot maxgen_{i} \cdot on_{t,i} + startcost_i\cdot start_{t,i}\right) }{}{} 
	\breakObjective{+A\cdot\sum_{t=1}^T\left(\sum_{i=1}^I maxgen_{i}\cdot on_{t,i}-rd_t\right)^2}
	\breakObjective{+B\cdot\sum_{t=1}^T \sum_{i=1}^I \left(on_{t,i}\cdot(1-on_{t-1,i}) + 2\cdot start_{t,i}\cdot\left(on_{t-1,i}+1-on_{t,i}\right)- start_{t,i}\right)}
	\breakObjective{+ C\cdot\sum_{t=1}^T \sum_{i=1}^I \left(start_{t,i}\cdot minup_i -\sum_{\tau=t}^{t-1+minup_i}start_{t,i}\cdot on_{\tau,i}\right)}
	\breakObjective{+ D\cdot\sum_{t=1}^T \sum_{i=1}^I \left(\sum_{\tau=t}^{t-1+mindown_i}\left(start_{t,i}+on_{t-1,i}-on_{t,i}\right)\cdot on_{\tau,i}\right)}
	\addConstraint{on_{t,i}, start_{t,i}}{\in\mathbb{B}}{\quad\forall~t=1,\ldots,T,~i=1,\ldots,I.}
\end{mini*}

\normalsize
The penalty terms are in the same order as the constraints for the classical formulation. By a closer look one can identify that if the $start$-variable is not properly set (i.e. it is zero although it has to be one), the last penalty term provides a bonus for an infeasible solution. Hence the penalty for wrongly setting $start$-variables has to be higher: $B>D\cdot\max_{i=1,\ldots,I} mindown_i$. It is also possible to model a UCP with energy generation between $mingen_i$ and $maxgen_i$ using discrete power generation steps.
Given a step size $step_i$, we need $d_i\coloneqq \lfloor\log_2\left(\nicefrac{(maxgen_i-mingen_i)}{step_i}\right)\rfloor + 1$ variables using a logarithmic encoding of the steps (\cite{lucas}). In total, our model needs $T\cdot I\cdot\left(2+\sum_{i=1}^I d_i\right)$ variables. Then we can replace the power output $maxgen_{i} \cdot on_{t,i}$ by
\scriptsize
\begin{equation*}
mingen_i\cdot on_{t,i} + \left(\sum_{k=1}^{d_i-1} 2^k\cdot gen_{t,i,k} + \left(d_i+1-2^{d_i}\right)\cdot gen_{t,i,d_i}\right)\cdot step_i.
\end{equation*}
\normalsize
Further, we have to add a coupling constraint between the $on$-variables and the discrete power generation variables $gen_{t,i,k}$:
\scriptsize
\begin{equation*}
	A\cdot\sum_{t=1}^T\left(\left(1-on_{t,i}\right)\cdot\sum_{k=1}^{d_i} gen_{t,i,k}\right).
\end{equation*} 
\normalsize

\section{Illustrative Example}\label{sec:example}
\begin{figure}[tb]
	\centering
	\includegraphics*[width=0.5\textwidth]{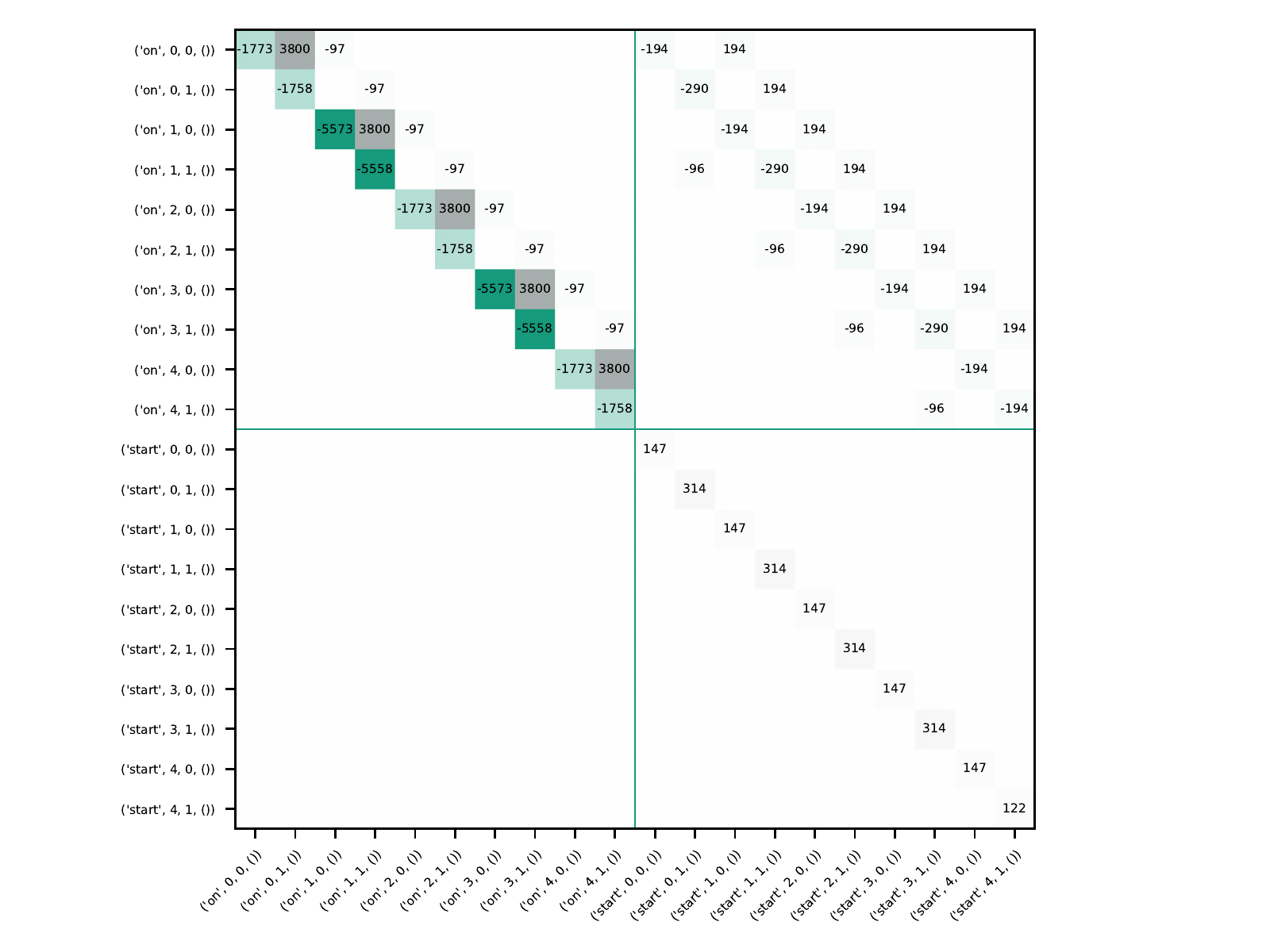}
	\caption{The QUBO matrix corresponding to our example. The first index counts time steps, the second the unit.}\label{fig:matrix}
\end{figure}

In order to illustrate our model, we have built the QUBO problem for an instance with two thermal units and five time steps. Further, we assume that each unit can only operate on maximal power generation if turned on. The parameter values are given in Table~\ref{tab:input}. Further, the residual demand is given by $\{1, 2, 1, 2, 1\}$. The upper triangular matrix of the QUBO problem with penalty factors $A=1900, B=97, C=96, D=96$ is visualized in Figure~\ref{fig:matrix}. 
\begin{table}[tb]
	\centering
\caption{Input data of our example.}\label{tab:input}
\scriptsize
\begin{tabular}[t]{l|l|l|l|l|l|l}
	Name   & $mingen_i$ & $maxgen_i$ &  $minup_i$ & $mindown_i$ & $startcost_i$ & $varcost_i$\\\hline
	Unit 0 &          1 &          1 &           1 &             1 &          50 &       30 \\
	Unit 1 &          1 &          1 &           2 &             1 &          25 &       45 \\
\end{tabular}
\normalsize
\end{table}

Clearly, we have obtained a sparse matrix, as only 55 of 400 entries ($13.75\%$) are nonzero. This number rapidly decreases to $1$ to $2\%$ for larger instances. Further, the maximum number of nonzero entries per row/column is $5$. Hence, corresponding qubits are less connected and the embedding onto real hardware needs only few extra qubits. In contrast, for the generic formulation, we get the following penalty term controlling the relation between $start$- and $on$-variables: $\left(start_{t, i} - (on_{t, i} - on_{t-1, i}+ slack_{t, i})\right)^2$. This results in an additional qubit for each $start$-variable with a necessary connection to the qubits responding to $start_{t, i}$,$on_{t, i}$, and $on_{t-1, i}$. The effect of one slack variable for every power plant and time step with only three interactions each may appear relatively harmless. Yet, for the minimum running time constraint this requires the introduction of at least $\lfloor\log_2(minup_i)\rfloor + 1$ slack variables for every unit and time step. Due to squaring, these interact with one $start$-variable and a number of $on$-variables each and also with each other. The same holds true for the minimum idle time.\footnote{Remark that in our example this parameter is set to 1 and does not need any binarization, increased parameter values worsen the generic model.} In total, the generic model has $50$ variables with $106$ interactions between different variables, while our tailor-made approach calls for only $38$ interactions of $20$ variables thus makes better use of system resources.

We have solved this model using the \emph{Gurobi} solver (Version 9.5.1) on classical hardware, the \emph{Quantum Approximate Optimization Algorithm (QAOA)} with warm start \cite{qoao} on the \emph{IBM Qiskit} 0.37.1 QASM simulator, and the \emph{D-Wave Quantum Annealer} (Version Advantage\_system 5.2 with over 5000 qubits) using 23 physical qubits instead of the 20 virtual qubits. The Gurobi solver used on the classical formulation finds the optimal solution $\{0, 1, 1, 1, 1, 0, 1, 1, 0, 1, 0, 1, 1, 0, 0,$ $0, 0, 1, 0, 0\}$ with objective value $370$ (Ising objective value -20530). Here, unit 0 is on in time steps 1 to 3. Unit 1 is running in time steps 0, 1, 3, and 4. For $1000$ shots, this is the solution occurring most often on D-Wave. The QASM simulator with warm started QAOA, however, does not find the optimal solution but finds $\{0, 1, 1, 1, 0, 1, 1, 1, 0, 1, 0, 1, 1, 0, 0, 0, 1, 0, 0, 0\}$ with objective value $410$ (Ising objective value -20490). This solution is feasible and just switches the occupancy of the units in time step 2, which is also reflected in changed states for both $start$ variables in time step 3.

\section{Conclusion}\label{sec:concl}
In this article, we have presented a new formulation for the unit commitment problem as a QUBO such that it can be solved using quantum computers. Our model correctly penalizes infeasible solution while providing a compact matrix with fewer necessary qubits and connections between qubits than generic translation methods. First tests show the advantages of our model in practice. While our model still has a binary encoding of the continuous variables occupying valuable qubits, a recent publication show that it may possible to encode these variables without binary encoding on some QC devices \cite{vqco}.

Besides the introduction of uncertainty to the residual demand, immediate future research will obviously be to carry out an extensive computational study of our model on various systems (e.g. Gurobi vs. quantum computer) comparing solution quality and running times. Due to the limited availability of qubits and connectivity in gate-based quantum computers, we will focus on the quantum annealers from D-Wave. This study promises to return answers to two important questions: can current quantum computing devices outperform classical solvers on classical hardware for practical problems and, if not, what is necessary in the future to ascertain \enquote{quantum supremacy} in practice?   

\paragraph{Acknowledgments}
This publication has been funded by the German Federal Ministry for Economic Affairs and Climate Action (grant no. 03EI1025A).

%
%

\end{document}